# Quantum Dynamics of Small-Size Triplet Superconductors with Broken Time-Reversal Symmetry[1]

A. M. Gulian[2] and K. S. Wood[3]

Abstract:

   Symmetry of the order parameter in some triplet superconductors corresponds to doubly-degenerate chiral states. We predict that in a sufficiently small sample this degeneracy can be lifted via macroscopic quantum tunneling. Moreover, if the initial state is prepared correspondingly, 'instanton'-like quantum behavior is possible in a single-domain sample. In this case *coherent oscillations* of magnetic momentum should take place, as well as *sign alternations* of the boundary currents. In particular, *spontaneous* Hall effect recently predicted for ruthenates should posses the oscillating voltage feature.

   It is plausible to attempt to achieve an oscillatory state in triplet superconductors with broken time-reversal symmetry; such complex phenomena can be anticipated by analogy with other quantum mechanical systems. At temperatures below the superconducting transition the state is doubly degenerate in these superconductors. In this article we show how the spin-triplet pairing creates the possibility of oscillation of states. We discuss the practical detectability of such oscillations and the relation of the oscillations to tunneling effects. These oscillations are predicted by analogy with the ammonia molecule and $^3$He-A, atomic Bose-Einstein condensates. The potential practical importance lies in applications of triplet state superconductivity as discussed in the concluding paragraph; there are applications where tunneling between and superposition of eigenstates is absolutely required. Since $Sr_2RuO_4$ [1-3] is regarded as a "textbook example" of spin-triplet pairing, we refer primarily to that material during this discussion. Other relevant candidates could be found among "heavy fermion" superconductors such as $UBe_{13}$ [4], as well as among quasi-one dimensional organic superconductors, such as the Bechgaard salts based on the tetramethyltetraselenafulvalene molecule, specifically $(TMTSF)_2PF_6$ [5].

   Competition between ferromagnetic and antiferromagnetic fluctuations in $Sr_2RuO_4$ [6] favors triplet pairing with total spin $S=1$ [7]. The order parameter in triplet superconductors can be presented as $\hat{\Delta} = i\mathbf{d}(\mathbf{k})\cdot\boldsymbol{\sigma}\sigma_y$, where $\boldsymbol{\sigma}=(\sigma_x, \sigma_y, \sigma_z)$ – are the Pauli matrices, and the vector $d_\alpha = A_{\alpha i}\hat{k}_i$ and $A_{\alpha i}$ is a 3×3-matrix of complex numbers (the indices $\alpha$ and $i$ correspond to the directions in spin and orbital (coordinate) space correspondingly, see, e.g., [8-10]). In the simplest case $A_{\alpha i} = \exp(i\Theta)\delta_{\alpha i}$. A corresponding analog is the pseudo-isotropic *(B-)* phase of $^3$He. The energy gap $|\Delta(\boldsymbol{\kappa})|=\Delta(T)$ and spin of Cooper-pairs has 3 projections: $S_z=1, 0,$ and $-1$. More intriguing are properties of anisotropic phases, such as the one corresponding to the so-called A-phase of $^3He$ [11]. In this case the energy gap has a nodal structure ($\Delta(\mathbf{k})=0$ at opposite poles on the Fermi-sphere. These poles specify a direction of anisotropy. The spin of Cooper pairs has only two non-zero projections: $S_x=1$ and $–1$ (Fig.1).

   $Sr_2RuO_4$ reveals a pronounced two-dimensional structure of the electron liquid (because conductivity along the crystallographic *c*-axis is more than *800* times less than conductivity

---

within the *(a,b)*-plane [1-3]). To describe the order parameter in ruthenates, one should analyze triplet superconductivity in two dimensions on a square lattice (in view of its tetragonal crystalline symmetry). Because of finite spin-orbital coupling, neither spin nor orbital moments are good quantum variables. Possible pairing states, including splitting due to the spin-orbit coupling, are usually classified by irreducible representations $\Gamma_1^- - \Gamma_5^-$ [7].

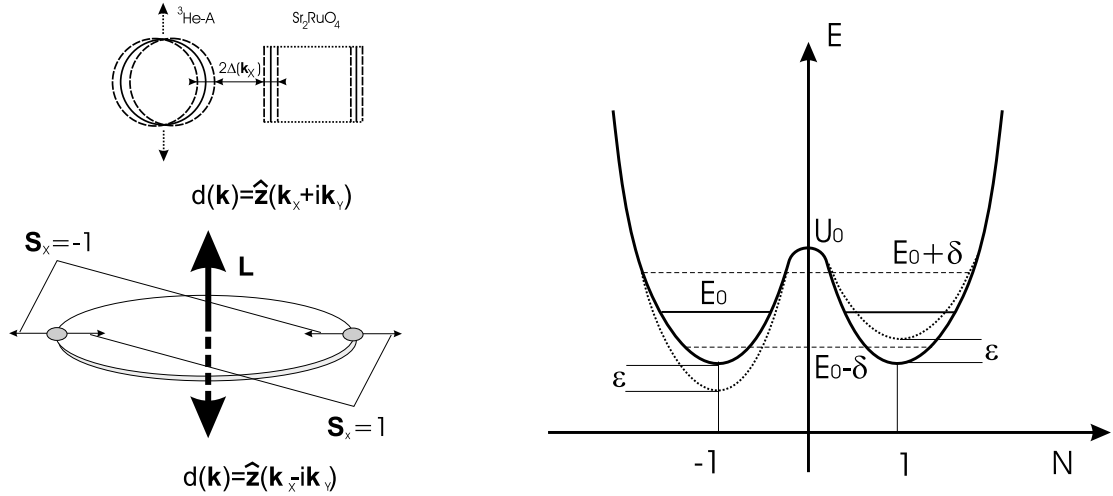

**Figure 1.**
A. (left top corner): Fermi-surface (solid lines) and energy gaps (dashed lines) in *³He-A* (circles, side view) and in *Sr₂RuO₄* (cylinders, also side view) and doubly-degenerate symmetry of Cooper pairs corresponding to the broken time-reversal symmetry (left bottom). Accordingly, the order parameter is also doubly degenerate.
B. (on the right). The double-well potential corresponding to doubly degenerate order parameter in the configuration space. 'N' denotes the chirality of the order parameter states. When the quantum wavefunction is collapsed, the qubit is either in state |-1> or |1>. This corresponds to the energy level $E_0$ in each of them. If the conditions are created facilitating quantum tunneling across the barrier, a transition into a coherent state can happen, so that the symmetric state $[2^{-1/2}(|-1> + |1>)]$ will have the energy $E_0-\Delta E$, and anti-symmetric state $[2^{-1/2}(|-1> - |1>)]$ will have the energy $E_0+\Delta E$. Transition between these two states should be experimentally detectable provided $\Delta E$ is more than the line-width of the level. Another, easier to detect, oscillatory effect between the states N=1 and N=-1 is possible according to quantum dynamics (see the text).

**Table 1. Some of possible triplet states in Ruthenate** [7].

In defining triplet states, available options include splitting due to spin-orbit coupling; x,y,z denote the directions in the spin–space. J and $J_z$ correspond to total angular momentum and its z-component respectively.

| $\Gamma$ | J, $J_z$ | d(k) |
|---|---|---|
| $\Gamma_1^-$ | 0,0 | $\hat{x}k_x + \hat{y}k_y$ |
| $\Gamma_2^-$ | 1,0 | $\hat{x}k_y - \hat{y}k_x$ |
| $\Gamma_3^-$ | 2,±2 | $\hat{x}k_x - \hat{y}k_y$ |
| $\Gamma_4^-$ | 2,±2 | $\hat{x}k_y + \hat{y}k_x$ |
| $\Gamma_5^-$ | 1,±1 | $\hat{z}(k_x \pm ik_y)$ |

The $\Gamma_5^-$ representation results in a symmetry for the wavefunction $\mathbf{d} = \hat{z}\Delta(k_x \pm ik_y)/k_F$ (with total angular momentum $J=1$, $J_z=\pm 1$ where $\hat{x}, \hat{y}, \hat{z}$ denote the directions in the spin-space) with $\hat{z}$ orthogonal to the conductivity plane. In the two-dimensional geometry this state is nodeless, but it is still analogous to the A-phase of $^3$He, since two polar nodes disappear at infinite stretching of the Fermi-surface (arrows on Fermi-sphere in Fig.1). It is very likely that this state which corresponds to the broken time-reversal symmetry (well confirmed by the muon experiments [12]) occurs in ruthenates.

Broken time-reversal symmetry means that the ground state of ruthenates should be doubly degenerate. C*hiral states* are a direct corollary of this observation. Chirality is characterized by a topological number $N = \frac{1}{4\pi}\iint dk_x dk_y \hat{\mathbf{m}} \cdot \left(\frac{\partial \hat{\mathbf{m}}}{\partial k_x} \times \frac{\partial \hat{\mathbf{m}}}{\partial k_y}\right)$, where the unit vector $\hat{\mathbf{m}}(k_x, k_y)$ is $\hat{\mathbf{m}} = \frac{\mathbf{m}}{|\mathbf{m}|}$, and where $\mathbf{m} = (\operatorname{Re} d_z, \operatorname{Im} d_z, \varepsilon_k)$, $\varepsilon_k = (k_x^2 + k_y^2)/2m - \mu$, $\mu$ being the chemical potential [13]. For the state $\Gamma_5^-$ : $N = \pm 1$, and in absence of external fields, *E(N=1)=E(N=-1)*. The duality of *N* implies possibility of multiple domains. As usual with broken symmetries [14], in each domain a certain value of N is established at cooling down of the superconductor because of fluctuating initial parameters, and it may be different from cooling to cooling, as in $^3$He [15]. Large samples will have multi-domain structure. Small enough samples should constitute a single domain. Regarding superconductors as quantum objects, one can expect having quantum superposition of states *|N=1>* and *|N=-1>* in the same domain: *|Σ>=α|1> + β|-1>*, so that *|α|²+|β|²=1*. For equal energy states this superposition is a consequence of quantum tunneling. To understand better the consequences let us analyze this idea in some details.

The 'spin-boson' Hamiltonian which corresponds to the situation described in Fig. 1 can be written in a form:

$$H_{SB} = E_0 \hat{I} - \delta \hat{\sigma} + \varepsilon \hat{\sigma} + H_{int} \tag{1}$$

where $\delta$ corresponds to tunneling, $\varepsilon$ is a 'tuning' parameter, an $H_{int}$ contains an interaction of tunneling system with the environment. We kept $E_0$ just for illustrative purposes, it can be dropped off at any point. Let us ignore for the moment the interaction with environment and put also $\varepsilon$=0. Then

$$\hat{H} = \begin{pmatrix} E_0 & -\delta \\ -\delta & E_0 \end{pmatrix} = E_0 \hat{I} - \delta \hat{\sigma}_x \tag{2}$$

The solution of Schrödinger's equation has then a form (we introduce $E_1=E_0-\delta$, $E_2=E_0+\delta$, and also put h/2π =1):

*|1>=(a/2)exp[-(iE₁t)] +(b/2)exp[-i(E₂)t]*

(3)

$|-1> = (a/2)\exp[-(iE_1 t)] - (b/2)\exp[-(iE_2 t)]$

If $b=0$ system is at a minimal energy state $E=E_1$ (symmetric configuration) and if $a=0$ – at a higher energy state $E=E_2$ (anti-symmetric configuration). These two states are the eigenstates of the Hamiltonian (2). If the system is in one of these states, it can stay there indefinitely long in absence of perturbations.

The situation is less static when the system initially is in a 'collapsed' state, either in $|1>$ or in $|-1>$. For that case, we should substitute $t=0$ in Eqs. (3), getting $a/2+b/2=1$, $a/2-b/2=0$, so that $a=b=1$. This immediately yields a well-known result: starting at $t>0$ the system will coherently oscillate between states $|1>$ and $|-1>$, so that corresponding probability difference $P(t) = P_{|1>} - P_{|-1>} = Cos(2\delta t)$ (quantum beating with the frequency $E_2-E_1$). Questions arise at this stage:

i) Does the oscillation-effect really exist?
ii) What is the expected range of oscillation frequencies?
iii) How does one detect this effect?
iv) What are other possible consequences of the coherent superposition of these chiral states?

The answer to question (*i*) is definitely 'yes' in an ideal case of environmentally decoupled system: $H_{int}=0$. As soon as $H_{int}\neq 0$ the frequency $\delta$ will be renormalized, and the damping will take place [16]. The role of damping can be negligible in some cases, and crucial in other cases, so that overdamping precludes any oscillation. To be able to make an unambiguous prediction a model described by tunneling Hamiltonian should correspond to reality. Tunneling between angular rotation states should be mapped on the tunneling due to mechanical motion. This problem is solved in the case of ferromagnetism [17]. The spectral function $J(\omega)$ characterizing $H_{int}$ [16] should also be specified. Another way to find an answer is to look into the closely related system: $^3$He-A, where coherent oscillations between chiral states are detected experimentally [15, 18]. Theoretical considerations based on analogy with 'instantons' in this quantum liquid reveal the possibility of coherent oscillations of the orbital moment between states **L** and –**L** [11, 19]. It is notorious that in this experiment the state must be prepared by switching on the magnetic field in specific directions. Oscillations of the orbital momentum begin as soon as the field is off. The height of the potential barrier $U_0$ (Fig. 1) has a direct influence onto the frequency: the period of oscillations rapidly becomes large when T drops far below $T_c$. The same should be expected in ruthenates, but one needs a special investigation to make quantitative predictions related to (ii).

An answer to question (iii) lies in the implications of the existence of chirality itself. In principle, the detection of oscillations is possible in different ways.

1. The oscillation of magnetic momentum of a small specimen: its amplitude is equal to [20]

$$|\mathbf{M}| = \left|\frac{1}{2c}\int d^3(\mathbf{r}\times\mathbf{j})\right| \approx \frac{e\hbar}{mc} n_S V \qquad (4)$$

(here $n_S$ is the density of Cooper pairs, $V$ is the specimen volume) and for a sample of an appropriate size this can be as large as $10^6$-$10^8$ Bohr's magnetons, which is a value detectable even at room temperature.

2. Surface currents (or boundary currents in case of films) are a direct implication of the *S=1* spin states of the Cooper pairs, since *S=1* requires orbital momentum *L=1*. These can be registered via the induction of counter-running currents in the conductors placed in the vicinity of these boundaries. Obviously, boundary currents should oscillate if system oscillates between chiral states.

3. One further method involves a spontaneous Hall effect, which was predicted in superconductors with broken time reversal symmetry [13]. The effect is related to the fact that the orbital motion of Cooper pairs yields surface currents without application of an external magnetic field, hence one can observe a Hall potential difference just because of the electric current flowing through the Hall bar. Instead of being proportional to magnetic field H, in this case $V_{Hall} \propto N$, the chiral number. Thus, applying a DC current one can register alternating $V_{Hall}$ because of the alternating N.

Having the critical temperature $T_c \sim 1.5K$, ruthenates have been regarded only as having academic interest, mainly useful for clarifying different aspects and mechanisms of superconductivity. Observation of macroscopic quantum tunneling in ruthenates can bring them to practical importance. As we mentioned in introduction, the order parameter of anisotropic triplet superconductors differs from a simple scalar form of ordinary superconductors. It has rather a vector character, which favors application of these unique objects for quantum information storage and processing [21]: one does not need any special circuitry to get qubits (the quantum computing analog of ordinary bits) thus avoiding additional sources of undesired decoherence, a pervasive problem in quantum computing [22].

**Acknowledgments**

The authors are grateful to H. Gursky, D. Van Vechten, M. Lovellette, P. Reynolds, Y. Liu, and M. Johnson for interest to this work and useful discussions. This work was supported in part by the U.S. Office of Naval Research under Grant N0001401WX21228.